\documentclass[12pt]{article}
\usepackage[dvips]{graphicx}
\begin{document}

\setlength{\parindent}{0pt}
\bibliographystyle{unsrt}
\title{Transition from  Quantum to Classical Information in a Superfluid}
\author{ A.Granik\thanks{Department of
Physics, University of the
Pacific,Stockton,CA.95211;~E-mail:~agranik@uop.edu}~~and~~G.Chapline\thanks{Lawrence
Livermore National Lab., Livermore,
CA.94550;E-mail:~Chapline1@LLNL.gov} }
\date{}
 \maketitle
\begin{abstract}
Whereas the  entropy of any deterministic classical system
described by a principle of least action is zero, one can assign a
"quantum information" to quantum mechanical degree of freedom
equal to Hausdorff area of the deviation from a classical path.
This raises the question whether superfluids carry quantum
information. We show that in general the transition from the
classical to quantum behavior depends on the probing length scale,
and occurs for microscopic length scales, except when the
interactions between the particles are very weak. This transition
explains why, on macroscopic length scales, physics is described
by classical equations.
\end{abstract}
\section{Introduction}
In this letter we consider the following two questions:\\

1) Can information be carried by a superfluid order parameter
$\psi$ in the same way that information is carried by , say, a
radio wave?\\

and\\

2) Since the superfluid order parameter $\psi$ depends on
entanglement {\it{a la}} Bogoluibov, is there any difference
between information carried by a superfluid order parameter and
quantum information?\\

The answer to the first question is pretty clearly yes, but the
answer to the second is not so obvious. It was shown by Bogoluibov
\cite{NB} that superfluids depend on existence of EPR-like
correlations between particles of opposite momenta. Since EPR
correlations play an important role in quantum computing
\cite{CB}, one might conjecture that spatial variations in the
order parameter qualify
as "quantum informations".\\

As was first pointed out by Planck \cite {MP}, the
Clausius-Boltzmann entropy of any classical system obeying a
principle of least action is zero. This serves as a hint that in a
search of the answer to the second question one should turn to the
path integral formulation of quantum mechanics where the principle
of least action is not valid anymore. One feature of the path
integral formulation which can serve our purposes is an existence
of quantum-mechanical paths \cite{FH} that are continuous but
non-differentiable everywhere.\\

Therefore a transition to a classical regime is characterized by
smoothing out the "irregularities" of a quantum-mechanical path.
This smoothing was ascribed to the process of averaging "over a
reasonable length of time to produce ... an 'average' velocity"
\cite{FH}. In fact, the emergence of a classical path is due to a
decreased resolution used in measuring the path's length.\\

This is clearly seen if one would use for the description of a
quantum-mechanical path the concept of Hausdorff length and
dimension \cite{LA}. It was found that independently of the length
definition, the Hausdorff dimension of a quantum-mechanical path
is $D_H=2$ as compared to the classical dimension $D_H=1$. The
transition from one regime to another can be demonstrated by
explicitly evaluating the Hausdorff length in \cite{LA} for any
value of the Hausdorff dimension.\

It turns out that for $$a)~ <\Delta
l>=\int_{{\cal{R}}_3}d^3x|{\bf{x}}||\psi_{\Delta
x}({\bf{x}},\Delta t)|^2$$ where $<\Delta l>$ is the average
distance a particle travels in a time $\Delta t$, its path length
is
\begin{equation}
\label{eq:1} <L>~ \sim (\overline{\Delta
x})^{D-1}\{\sqrt{\frac{\pi}{2}}\Phi(\frac{\sqrt 2(\overline{\Delta
x})} {\sqrt{1+4(\overline{\Delta x})^{2}}})
+\frac{\sqrt{1+4(\overline{\Delta x})^2}}{\overline{\Delta x}}
exp[-\frac{(\overline{\Delta x})^2} {\sqrt{1+4(\overline{\Delta
x})^2}}]\}
\end{equation}
Here $\Phi(y)=(2/\sqrt \pi)\int_0^y e^{-y^2}$, $\overline{\Delta
x}=\Delta x/\lambda$,  $\Delta x$ is the spatial resolution,
$\lambda =\hbar/p_{av}$ is the de Broglie wavelength and $p_{av}$
is particle's average momentum. Fig.1 illustrates the dependence
given by (\ref{eq:1})\\
\begin{figure}
 \begin{center}
\includegraphics[width=6cm, height=6cm]{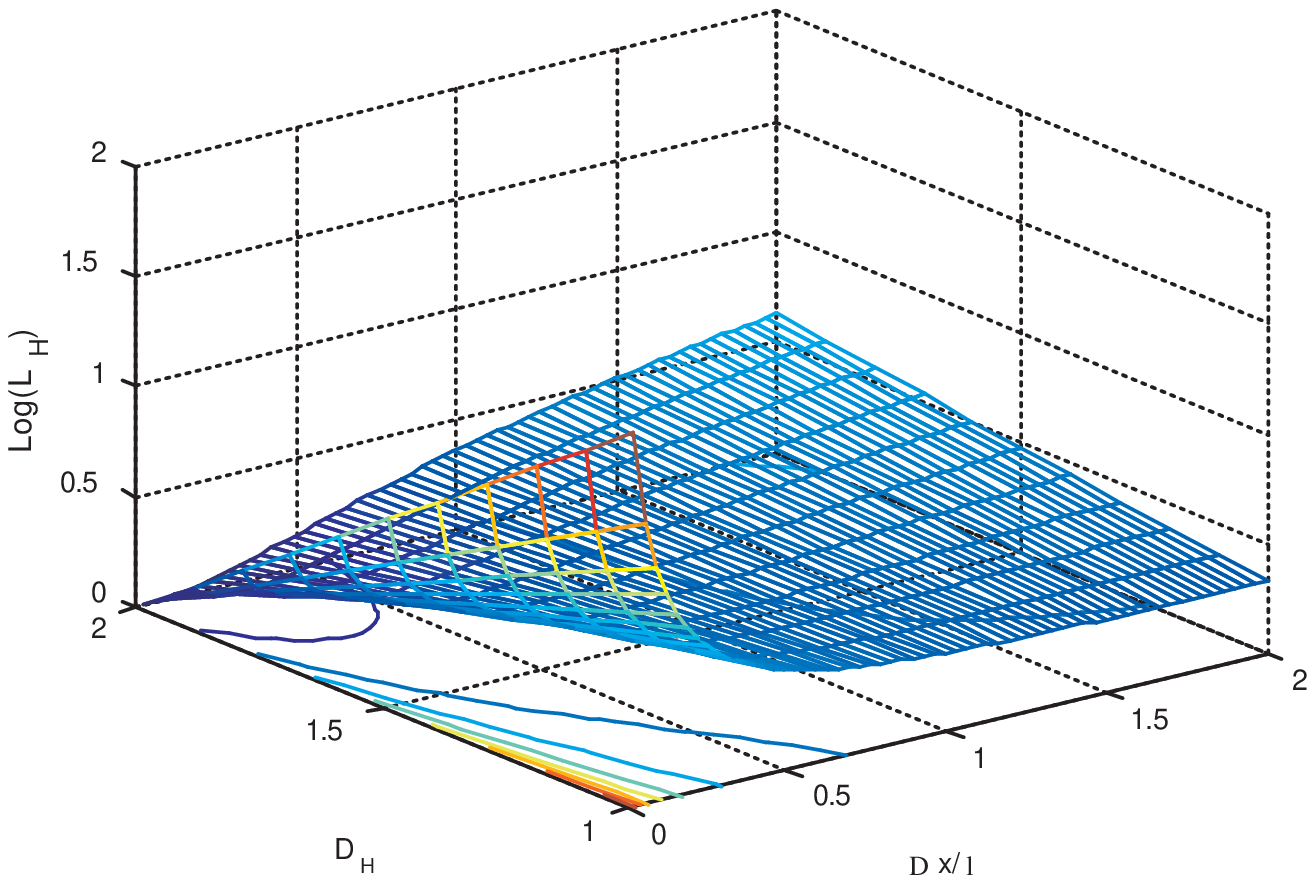}
\caption{\small  Log of Hausdorff length according to Eq.(1)as a
function of the Hausdorff dimension $D_H$ and the resolution
$\Delta x$ for Abbot-Wise analysis  \cite{LA} with $f(|{\bf
k}|)=exp(-|{\bf k}|^2)$}
 \end{center}
 \end{figure}

If we use another definition of length $$b) <\Delta l>
=\sqrt{\int_{{\cal{R}}_3}d^3x|{\bf{x}}|^2|\psi_{\Delta
x}({\bf{x}},\Delta t)|^2}$$ then the resulting path length is
\begin{equation}
\label{eq:2} <L> ~\sim ({\overline{\Delta
x}})^{D-2}\sqrt{16({\overline{\Delta x}})^2+3}
\end{equation}
The respective graph is shown in Fig.2.\\
\begin{figure}
 \begin{center}
\includegraphics[width=6cm, height=6cm]{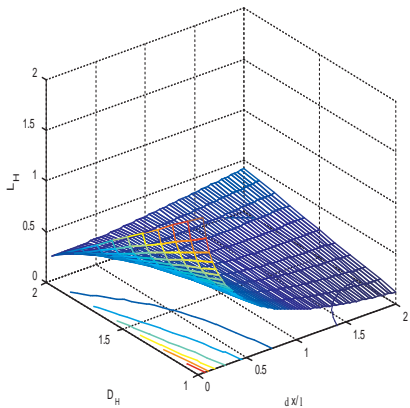}
\caption{\small  Log of Hausdorff length according to Eq.(2) as a
function of the Hausdorff dimension $D_H$ and the resolution
$\Delta x$ for Abbot-Wise analysis  \cite{LA} with $f(|{\bf
k}|)=exp(-|{\bf k}|^2)$;}
 \end{center}
 \end{figure}

In both cases the parameter describing a transition from a
classical to quantum regime ( and vice versa) is the dimensionless
spatial resolution
\begin{equation}
\label{eq:3} \overline{\Delta x}=\frac{\Delta x}{\hbar/p_{av}}
\end{equation}
It has a suggestive physical meaning: a ratio of a resolution
(physically implementable by some measuring classical device) used
to measure length (a probing scale) and the "characteristic"
length scale intrinsic to quantum process, which can be viewed as
De Broglie wave length. In that we made full circle on a spiral,
referring to the earlier view of quantum mechanics, but this time
on a higher level. Roughly speaking, the magnitude of this
parameter indicates how strong(weak) are quantum effects as viewed
from the classical world. As will be seen below, this parameter
has a universal character, and it will emerge in our discussion of
the
\section{Bose Condensate and the transition from Classical to
Quantum Behavior}

To illustrate this point we consider a Bose condensate of
interacting bosons at zero temperature \cite {GC}. Small
perturbations to the superfluid order parameter satisfy the linear
Schroedinger-like equation:
\begin{equation}
\label{eq:4} \frac{\partial^2}{\partial t^2}\phi
=v_s^2\nabla^2\phi-(\frac{\hbar}{2M})^2\nabla^4 \phi
\end{equation}
where $v_s$ is the speed of sound and $M$ is the mass of the
fluid.\\

We will consider the classical- quantum transition  with the help
of $2$ methods which , as will be seen later, turn out to be equivalent:\\

{\bf i) Numerical Calculation }\

In this approach the average distance $<\Delta l>$ the particle
travels in time $\Delta t$ can be written as follows:
\begin{equation}
\label{eq:5} <\Delta l>=\int_{{\cal R}^3}d^3x|{\bf x}||\Psi({\bf
x},\Delta t)|^2
\end{equation}
Here
\begin{equation}
\label{eq:6} \Psi({\bf x},\Delta t)=\frac{(\Delta
x)^{3/2}}{\hbar^3}\int_{{\cal
R}^3}\frac{d^3p}{(2\pi)^{3/2}}f(\frac{|{\bf p}|\Delta
x}{\hbar})e^{i{\bf p \bullet x}/\hbar-E\Delta t/\hbar}
\end{equation}
and $E$ is given in \cite{GC} as
\begin{equation}
\label{eq:7} E=\sqrt{(pv_s)^2+\frac{p^4}{4M^2}}
\end{equation}
By introducing the dimensionless quantities
\begin{equation}
\label{eq:A} {\bf k}=\frac{{\bf p}\Delta
x}{\hbar},~~\overline{\Delta x}=\Delta x\frac{Mv_s}{\hbar},~~{\bf
y}=\frac{{\bf x}}{\Delta x},~~\alpha=\frac{Mv_s^2\Delta
t}{\hbar}
\end{equation}
 and using (\ref{eq:7}) and (\ref{eq:6}) we rewrite the
expression (\ref{eq:5}):
\begin{equation}
\label{eq:8} <\Delta l> = \Delta x\int_{{\cal R}^3}d^3y|{\bf
y}||\int_{{\cal R}^3}d^3k f(k)e^{i{\bf k}\bullet{\bf
y}-i\alpha\sqrt{(k/\overline{\Delta x
})^2+\frac{1}{4}(k/\overline{\Delta x })^4}}|^2
\end{equation}
Interestingly enough, the dimensionless quantities $\alpha$ and
$\overline{\Delta x}$ have a very simple physical meaning:
$\overline{\Delta x}$ is the ratio of the classical and quantum
momenta (here the classical quantity appears naturally and not
introduced by hand as in \cite {LA}) and $\alpha$ is the ratio of
the respective energies.\\

If we take the function $f|{\bf k}|$ to be Gaussian , that is
$f|{\bf k}|=e^{-|{\bf k}|^2}$ then (\ref{eq:8}) yields the
following Hausdorff length $<L>$
\begin{equation}
\label{eq:9} <L>\sim {\overline{\Delta x}}^D\int_{{\cal
R}^3}d^3y|{\bf y}||\int_{{\cal R}^3}d^3k e^{-|{\bf k}|^2+i{\bf
k}\bullet{\bf y}-i\alpha\sqrt{(k/\overline{\Delta x
})^2+\frac{1}{4}(k/{\overline{\Delta x}})^4}}|^2
\end{equation}\\

In 2 limiting cases of\\

1) a purely classical regime ($\overline{\Delta x} \gg k$)\

 and\

2) a purely quantum regime ($ \overline{\Delta x}\ll k$)\\

the above expression can be evaluated analytically. For simplicity
sake ( and without any loss of generality) we consider a 1-D
realization of (\ref{eq:9}).\\

$1$) After some algebra, the first case yields the following
expression for the Hausdorff length $<L>$:
\begin{equation}
\label{eq:10}
 <L>\sim (\Delta
x)^D\{\sqrt{\pi}\frac{v_s}{v_{av}}\Phi(\frac{1}{\sqrt
2}\frac{v_s}{v_{av}})
 -2e^{-\frac{1}{2}(v_s/v_{av})^2}\}
 \end{equation}
where we denote $v_{av}=\Delta x/\Delta t.$ The result is rather
trivial, since in this case in the limit of $\Delta x\rightarrow
0$ the Hausdorff length is the conventional length, whose dimension is $D=1$.\\

$2)$ Quite analogously we find that in this case the Hausdorff
length is
\begin{equation}
\label{eq:11} <L>\sim \overline{\Delta x}^D\sqrt{1+(\frac{\hbar
\Delta t }{M\overline{\Delta x}^2})^2}
\end{equation}
In the limit $\overline{\Delta x}\rightarrow 0$ Eq.(\ref{eq:11})
yields
\begin{equation}
\label{eq:12} L\sim{\Delta x}^{D-2}
\end{equation}
which is the same result as was obtained for quantum case in
\cite{LA}. \\
\begin{figure}
 \begin{center}
\includegraphics[width=6cm, height=6cm]{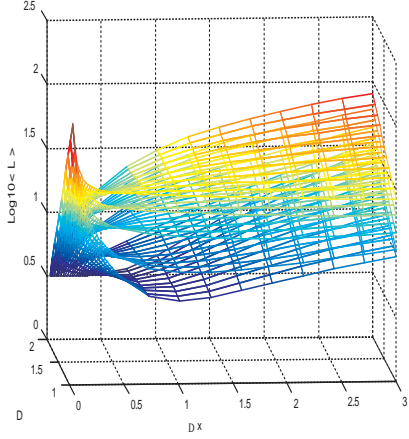}
\caption{\small  The  Hausdorff length according to Eq.(9) as a
function of the Hausdorff dimension $D_H$ , resolution $\Delta x,$
and sound speed $v_s=0,2,4,6,8$(bottom to top)}
 \end{center}
 \end{figure}\\

In general, the integrals in Eq.( \ref{eq:9}) cannot be found in a
closed form. Instead we evaluate them numerically. The result is
shown in  Fig.$3$ where the Hausdorff length is presented as a
function of the Hausdorff dimension $D_H$ , resolution $\Delta x$
and the following values of sound speed $v_s=0,2,4,6,8.$  \\

{\bf ii)Heuristic(De Broglie) Construction}  \\

To compare the results of the previous section with the
calculations based on an heuristic( De Broglie) picture, we begin
with the dispersion relation which follows from (\ref{eq:4}):
\begin{equation}
\label{eq:13} \omega_k=k v_s\sqrt{1+(\frac{\hbar k}{2Mv_s})^2}
\end{equation}
where $k=2\pi/$$\Lambda$ and $\Lambda$  is the wave length of the
small perturbations which we take as the probing length
scale,that is $\Delta x=\Lambda$. \\

In the De Broglie picture the group velocity is considered as a
velocity of a quantum "particle" which in turn would allow us to
introduce an analogue of the path length $<L>$ travelled by such a
particle in a time $T$. Moreover, since we are dealing with the
quantum path, this path length must be understood in Hausdorff
sense:
$$<L_H>=(\Delta x)^{D-1}<L>$$

From (\ref{eq:13}) we find the group (particle) velocity $v_g$
\begin{equation}
\label{eq:14} v_g=v_s\frac{2+(\frac{\hbar/Mv_s}{\Delta x
})^2}{2\sqrt{1+(\frac{\hbar/Mv_s}{2\Delta x})^2}}
\end{equation}\\
The path length $<l>$ (in its conventional sense) travelled by
this particle in a time interval $T$ is then
\begin{equation}
\label{eq:15} <l>=v_gT=Tv_s\frac{2+(\frac{\hbar/Mv_s}{\Delta x
})^2}{2\sqrt{1+(\frac{\hbar/Mv_s}{2\Delta x})^2}}
\end{equation}
We notice that in this case the dimensionless quantity
\begin{equation}
\label{eq:B} \overline{\Delta x} \equiv \frac{Mv_s\Delta x}{\hbar}
\end{equation}
emerges which is exactly the same as in the previous (numerical)
case (cf. Eq.\ref{eq:A}). As a result we obtain from (\ref{eq:15})
the following expression for the Hausdorff length $<L>_H$:
\begin{equation}
\label{eq:16} <L>_H\sim (\Delta x)^{D-2}\frac{1+2(\overline{\Delta
x})^2} {\sqrt{1+(2\overline{\Delta x})^2}}
\end{equation}
From Eq.(\ref{eq:16}) follows that for the limiting case of
$v_s=0$ (that is purely quantum case) the Hausdorff dimension is
$D=2$. On the other hand, for another limiting case of
$\overline{\Delta x}\gg 1$ , that is $Mv_s \gg \hbar/\Delta x$ the
Hausdorff dimension is $D=1,$ corresponding to the classical
limit. \\

The graph of the general dependence
$Log(<L_H>)=f(D,\overline{\Delta x},\overline{v_s})$ according to
Eq.(\ref{eq:16}), that is according to De Broglie picture, is
shown in Fig.4.
\begin{figure}
 \begin{center}
\includegraphics[width=6cm, height=6cm]{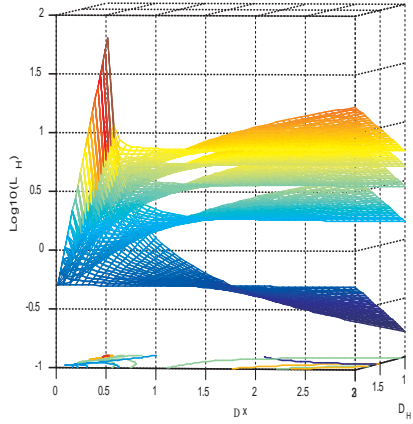}
\caption{\small  Log of the dimensionless Hausdorff length
according to Eq.(16) as a function of the Hausdorff dimension
$D_H$ , $\Delta x,$  and sound speed $v_s=0,2,4,6,8$(bottom to
top) according to the de Broglie picture}
 \end{center}
 \end{figure}
A comparison of Figs.(3) and (4) shows a remarkable qualitative
similarity between the Hausdorff length $<L>_H$ calculated on the
basis of the De Broglie picture and the same length found with the
help of the numerical analysis. However, this should not be very
surprising, if we take into account that the integrand in
Eq.(\ref{eq:9}) contains as a power of the exponent the dispersion
relation (\ref{eq:5}).\

It is seen (both from Eq.\ref{eq:16} and Figs.$3$ and $4$) that
the transition from quantum to classical regime ( characterized by
a change of Hausdorff dimension) is a continuous process, such
that a change from a classical to quantum regime is governed by
the dimensionless parameter (dimensionless length)
$\overline{\Delta x} ($ Eq.\ref{eq:B}) whose physical meaning was
given earlier as the ratio of the quantum "momentum" ( in the De
Broglie sense) and the classical momentum of a particle moving
with the speed of sound. A gradual increase of $\overline{\Delta
x} $ signals a continuous transition from a purely classical
($\overline{\Delta x} \rightarrow \infty$) to purely quantum
($\overline{\Delta x}=0$) regime.\
\section{Conclusion}

The emerging picture allows us to answer the second question posed
at the beginning of this letter. For small perturbations a
superfluid order parameter $\psi$ is a function of the parameter
$\overline{\Delta x}$. On the other hand, a continuous change of
this parameter  from $\overline{\Delta x}\rightarrow \infty$ to
$\overline{\Delta x}=0$ describes a transition from the classical
regime in a superfluid to the quantum regime. Therefore in the
limit $\overline{\Delta x}\rightarrow 0$ there is no difference
between the information carried by the superfluid order parameter
and the quantum information.\\

Amazingly enough, the above process describes a continuous
transition from a fluid-like coherent state to a coherent state of
weakly interacting particles. For ordinary sound velocities this
transition occurs for microscopic probing length scales. However,
if we are near a quantum critical point where $v_s\rightarrow 0$,
then this transition will occur for
macroscopic length scales.\\

Our results run contrary to the conventional point of view
regarding the transition from classical to quantum physics as
being necessarily due to decoherence \cite{WZ}. Indeed, in our
view decoherence plays essentially no role in the transition from
ordinary  classical physics to quantum physics. To the contrary
our results strongly support  the view \cite{GC} that the validity
of classical equations of motion for macroscopic length scales is
a consequence of having a vacuum state with a 'stiff' order
parameter. In fact, identifying ordinary spacetime with a
superfluid-like quantum state  with a small value of $\hbar/M v_s$
would be a natural result in almost any physically reasonable
quantum theory of gravity \cite{gc1}

\section{\bf Acknowledgements}
The authors would like to thank V.Panico for his valuable help in
preparation of this paper.\\

\section{Addendum}
After completion of this paper we became aware of a recent paper
by  Y. Shi \cite{YS}, that also shows that the transition from
quantum to classical behavior in a superfluid is not due to
decoherence, but using rather different arguments.

\end{document}